\documentclass[english,a4paper,11pt]{article}

\usepackage[english]{babel}
\usepackage{amsmath, amsxtra, amsfonts, amssymb, amstext}
\usepackage{amsthm}
\usepackage{booktabs}
\usepackage{fullpage}
\usepackage{nicefrac}
\usepackage{xspace}
\usepackage[noadjust]{cite}
\usepackage{url}
\usepackage{graphics}
\usepackage[usenames,dvipsnames]{xcolor}
\usepackage[colorlinks]{hyperref}
\definecolor{linkblue}{rgb}{0.1,0.1,0.8}
\hypersetup{colorlinks=true,linkcolor=linkblue,filecolor=linkblue,urlcolor=linkblue,citecolor=linkblue}
\usepackage[algo2e,ruled,vlined,linesnumbered]{algorithm2e}
\newtheorem{theorem}{Theorem}

\newcommand{\R}{\mathbb{R}}

\renewcommand{\epsilon}{\varepsilon}

\DeclareMathOperator{\E}{E}
\DeclareMathOperator{\Var}{Var}

\newcommand{\assign}{\leftarrow}

\newcommand{\onemax}{\textsc{OneMax}\xspace}

\begin{document}
    
\title{Collecting Coupons with Random Initial Stake}% or Very Precise Runtime Estimates for Randomized Local Search on Monotonic Functions}

\author{Benjamin Doerr$^{1,2}$,
Carola Doerr$^{2,3}$}
\date{$^1$\'Ecole Polytechnique, Palaiseau, France\\
$^2$Max Planck Institute for Informatics, Saarbr\"ucken, Germany\\
$^3$Universit\'e Paris Diderot - Paris 7, LIAFA, Paris, France
\\[2ex]
\today
}

\maketitle

\begin{abstract}
Motivated by a problem in the theory of randomized search heuristics, we give a very precise analysis for the coupon collector problem where the collector starts with a random set of coupons (chosen uniformly from all sets). 

We show that the expected number of rounds until we have a coupon of each type is $nH_{n/2} - 1/2 \pm o(1)$, where $H_{n/2}$ denotes the $(n/2)$th harmonic number when $n$ is even, and $H_{n/2}:= (1/2) H_{\lfloor n/2 \rfloor} + (1/2) H_{\lceil n/2 \rceil}$ when $n$ is odd. Consequently, the coupon collector with random initial stake is by half a round faster than the one starting with exactly $n/2$ coupons (apart from additive $o(1)$ terms).

This result implies that classic simple heuristic called \emph{randomized local search} needs an expected number of $nH_{n/2} - 1/2 \pm o(1)$ iterations to find the optimum of any monotonic function defined on bit-strings of length $n$.
%\merk{Abstract auch in der Coupon Collecotr Sprache?}
%We provide a very precise bound for the expected runtime of the randomized local search heuristic optimizing monotonic functions. This problem is equivalent to a variant of coupon collector problem where the collector starts with a random set of coupons (chosen uniformly from all sets). 
%
%We show that the expected number of rounds needed to solve this problem is $nH_{n/2} - 1/2 \pm o(1)$, where $H_{n/2}$ denotes the $(n/2)$th harmonic number when $n$ is even, and $H_{n/2}:= 1/2 H_{\lfloor n/2 \rfloor} + 1/2 H_{\lceil n/2 \rceil}$ when $n$ is odd.
\end{abstract}
%\textbf{Category:} {F.2.2}{Theory of Computation}{Analysis of Algorithms and Problem Complexity}[Nonnumerical Algorithms and Problems]\\


\sloppy{
\section{Introduction}
The coupon collector problem ask for how many coupons, chosen independently at random from $n$ different types, one needs to draw until one has a coupon of each type. It is well known and easy to prove that this number equals 
$n H_n$, where $H_n$ denotes the $n$th harmonic number $H_n:=1+\frac12+\frac13+\ldots+\frac1n$. The harmonic numbers are well understood. We have $H_n = \ln n + \gamma + \frac{1}{2n} \pm \Theta(n^{-2})$, where $\gamma \approx 0.5772156649\ldots$ is the Euler-Mascheroni constant. Consequently, the coupon collector needs $n \ln n + \gamma n + 1/2 \pm o(1)$ rounds to finish. 

If we assume that we start the coupon collector process being already in possession of $k$ different types of coupons, then the expected number of rounds becomes $n H_{n-k}$, and again the precise estimates for the harmonic numbers give very precise values for these numbers. 

Motivated by works in the theory of randomized search heuristics (see the following section), in this note we regard the coupon collector problem where we start with a random set of different coupon types. In other words, at the start of the process we already have each type with probability $1/2$, mutually independent among all types. Since in expectation we start with $n/2$ different coupons and the actual number of different coupons is strongly concentrated around this expectation, it seems natural that this coupon collector process should need roughly $n H_{n/2}$ rounds, that is, roughly the same time as if we would assume that we start with exactly $n/2$ different coupons. In fact, this was shown recently, however, the bound is less precise than the ones stated for the original coupon collector. In~\cite{Witt13}, Witt shows that the coupon collector process with random initial setting takes an expected number of $n H_{n/2} \pm o(n)$ rounds to finish.

In this work, we derive a bound sharp up to additive $o(1)$ terms. Interestingly, the answer is by an additive term of $1/2$ lower than $n H_{n/2}$.

\begin{theorem}\label{thm:randomcoupon}
  The coupon collector starting with a random set of different coupons takes $n H_{n/2} - 1/2 \pm o(1)= n \ln (n/2) + \gamma n +1/2 \pm o(1)$ 
  rounds to finish, where $H_{n/2}$ is the $(n/2)$th harmonic number when $n$ is even, and $H_{n/2}:= \frac12 H_{\lfloor n/2 \rfloor} + \frac12 H_{\lceil n/2 \rceil}$ when $n$ is odd.
\end{theorem}

Before proving the theorem, we briefly describe the motivation for this coupon collector variant stemming from the theory of randomized search heuristics.

\section{Optimizing Monotonic Functions via Randomized Local Search}

Randomized Local Search (RLS) is a simple hill-climbing heuristic. When used to maximize pseudo-Boolean functions $f: \{0,1\}^n \to \R$, RLS starts with a random initial search point $x \in \{0,1\}^n$. In each iteration it creates from $x$ a new search point (\emph{offspring}) $y$ by choosing a position $j \in [n]:=\{1,2,\ldots, n\}$ uniformly at random and flipping in $x$ the entry in the $j$th position. 
It replaces $x$ by $y$ if and only if $f(y) \geq f(x)$, i.e., if and only if the objective value (\emph{fitness}) of $y$ is at least as large as the fitness of $x$. 

\begin{algorithm2e}%
	\textbf{Initialization:} 
	Sample $x \in \{0,1\}^n$ uniformly at random\;
 \textbf{Optimization:}
\For{$t=1,2,3,\ldots$}{
Choose $j \in [n]$ uniformly at random\;
Set $y\assign x\oplus e^n_{j}$\,; //mutation step\\
\lIf{$f(y)\geq f(x)$}{$x \assign y$\,; //selection step}
}
\caption{Randomized Local Search for maximizing~$f\colon\{0,1\}^n\to\mathbb{R}$.}
\label{alg:RLS}
\end{algorithm2e}

A pseudo-Boolean function $f: \{0,1\} \to \R$ is called (strictly) \emph{monotonic} if for all $x, y \in \{0,1\}^n$, we have $f(x) < f(y)$ whenever $x \le y$ (point-wise) and $x \neq y$. It is easy to see that RLS optimizes all monotonic functions in an identical fashion. Since bits in $x$ that are one never change to zero, we in fact simulate a coupon collector process with random initial state. Since the particular monotonic function is not important,  often only the simple \onemax function $f : x \mapsto \sum_{i \in [n]} x_i$, counting the number of ones, is regarded. Summarizing this discussion, Theorem~\ref{thm:randomcoupon} is equivalent to the following statement.

\begin{theorem}
\label{thm:main}
The expected number of iterations until RLS on \onemax or any other monotonic function generates an optimal search point is 
$n H_{n/2} - 1/2 \pm o(1)= n \ln (n/2) + \gamma n +1/2 \pm o(1)$.
\end{theorem}

\section{Proof of the Main Result}

We first give the arguments for even values of $n$. The case of odd $n$ will be dealt with at the end of this section.

\paragraph{Even values of $n$.}
Let $X$ be the random variable that describes the number of ones in the initial search point. $X$ is binomially distributed with parameters $n$ and success probability $1/2$.

The expected number of iterations until RLS on \onemax generates an optimal search point can be written as
\begin{align}
\label{eq:basic}
\E[T] = \sum_{i=0}^n{E[T \mid X = i] \Pr[X=i]}\,.
\end{align}
By the definition of binomial distributions we have $\Pr[X=n/2-a] = \Pr[X=n/2+a]$ for all $a \in [n/2]$.
The main idea of this proof is to show that the average runtime 
$1/2 \left( \E[T \mid X = n/2+a] + \E[T \mid X = n/2-a] \right)$ does not deviate from 
$\E[T \mid X = n/2]$ by much.
Let $\varepsilon_a:=2 \E[T \mid X = n/2] - \left( \E[T \mid X = n/2+a] + \E[T \mid X = n/2-a] \right)$. 
We can rewrite (\ref{eq:basic}) to 
\begin{align}
\label{eq:basic2}
\E[T] = \E[T \mid X=n/2] - \sum_{a=1}^{n/2}{\Pr[X=n/2+a] \varepsilon_a}\,.
\end{align}

Consequently, it suffices to show that 
\begin{align*}
%\label{error1}
d:= \sum_{a=1}^{n/2}{\Pr[X=n/2+a] \varepsilon_a} = 1/2 \pm o(1)\,.
\end{align*}
Using the already mentioned fact that
\begin{align*}
E[T \mid X = k] =\sum_{i=k}^{n-1}{n/(n-i)}= n \sum_{i=1}^{n-k}{1/i}
\end{align*}
we compute 
\begin{align}
\varepsilon_a 
& = 
	2 E[T \mid X = n/2] 
	- 
	\left( \E[T \mid X = n/2+a] + \E[T \mid X = n/2-a]\right)
\nonumber
\\ & =
	2n \sum_{i=1}^{n/2}{1/i} -
	\left( 
	n \sum_{i=1}^{n/2-a}{1/i} + n \sum_{i=1}^{n/2+a}{1/i}
	\right) 
\nonumber	
\\& =
	n \left( \sum_{i=n/2-a+1}^{n/2}{1/i} - \sum_{i=n/2+1}^{n/2+a}{1/i} \right)
\nonumber	
\\ & =
	n \sum_{i=0}^{a-1}{\left(\frac{1}{n/2-i} - \frac{1}{n/2+i+1}\right)}
\nonumber	
\\ & = 
	n \sum_{i=0}^{a-1}{\frac{2i+1}{(n/2-i)(n/2+i+1)}}
\label{eq:epsa}	
\end{align}

Consequently,
\begin{align*}	
\varepsilon_a 
	\geq 
	 n \sum_{i=0}^{a-1}{\frac{2i+1}{(n/2)^2+n/2}}
=	
	\frac{4 a^2}{n+2} 
	\,
\end{align*}
and thus $d
 \geq 
\sum_{a=1}^{n/2}{\Pr[X=n/2+a] \frac{4 a^2}{n+2}}$.
Fortunately, 
\begin{align}
\label{eq:var}
\sum_{a=1}^{n/2}{2 a^2 \Pr[X = n/2 + a]} = \E[(X-\E[X])^2] =: \Var[X] \,
\end{align}
is just the variance of $X$, which is known to equal $n/4$, so we easily obtain the lower bound
\begin{align*}
d & \geq
 \frac{2 \Var[X]}{n+2}
 =
 \frac{n}{2(n+2)} = 
 1/2 -o(1)
\,.
\end{align*}

To bound the deviation $d$ from above, using again  (\ref{eq:epsa}) we estimate $\varepsilon_a $ by
\begin{align*}
\varepsilon_a 
& = 
	n \sum_{i=0}^{a-1}{\frac{2i+1}{(n/2-i)(n/2+i+1)}}
	\\&  \leq 
	n \sum_{i=0}^{a-1}{\frac{2i+1}{(n/2)^2-i^2}}
	\\ & \leq
	n \sum_{i=0}^{a-1}{\frac{2i+1}{(n/2)^2-(a-1)^2}}
	\\ & = 
	 \frac{n a^2}{n^2/4 - (a-1)^2}\,.
\end{align*}

Setting $A:=\sqrt{c n \ln n}$ for a constant $c$ to be fixed later, we split 
\begin{align}
d \label{eq:Aterm}
\leq & 
\sum_{a=1}^{n/2}{\Pr[X=n/2+a] \frac{n a^2}{n^2/4 - (a-1)^2}}
\\
 \leq &
\nonumber
\sum_{a=1}^{A}{\Pr[X=n/2+a] \frac{n a^2}{n^2/4 - (A-1)^2}}\\
\nonumber	& - 
	 \Pr[X-n/2 > A] \frac{n (n/2)^2}{n^2/4 - (n/2-1)^2}
		\,.
\end{align}

We first bound the last term in this expression.
Using Chernoff's bound (see, e.g.,~\cite{Doerr11bookchapter}), it is easy to see that 
\begin{align*}
\Pr[X-n/2 > A] \leq \exp(-2c \ln(n)) = 1/n^{2c}\,.
\end{align*}
Furthermore, 
\begin{align*}
\frac{n (n/2)^2}{n^2/4 - (n/2-1)^2} \leq n^3\,.
\end{align*}
Thus, for $c>3/2$ the last term in (\ref{eq:Aterm}) is clearly $o(1)$.

For bounding the other terms we use again fact (\ref{eq:var}) and the equality $\Var[X]=n/4$ to obtain
\begin{align*}
\sum_{a=1}^{A}{\Pr[X=n/2+a] \frac{n a^2}{n^2/4 - (A-1)^2}}
& \leq 
\frac{n \Var[X]}{2(n^2/4 - (A-1)^2)}
\\
& =
\frac{n^2}{2n^2 - 8(A-1)^2}
= 1/2+o(1)
\,.
\end{align*}
Plugging this back into (\ref{eq:Aterm}), we obtain the upper bound
\begin{align*}
d
\leq 
1/2+o(1)
\,.
\end{align*}

\paragraph{Odd values of $n$.}
It remains to consider odd values of $n$. The computations are very similar to the ones above. 
We keep the notation from the proof above, but we set, for $1 \leq a \leq \lfloor n/2 \rfloor$
\begin{align*}
\varepsilon_a:= \E[T \mid X = \lfloor n/2 \rfloor] + \E[T \mid X = \lceil n/2 \rceil]  - \left( \E[T \mid X = \lfloor n/2 \rfloor-a] + \E[T \mid X = \lceil n/2 \rceil+a] \right)\,.
\end{align*} 

Similarly to (\ref{eq:basic2}) we have 
\begin{align*}
%\label{eq:basic3}
\E[T] =  
1/2\left(
\E[T \mid X = \lceil n/2 \rceil] + \E[T \mid X = \lfloor n/2 \rfloor] 
\right) 
- \sum_{a=1}^{\lfloor n/2 \rfloor}{\Pr[X=\lfloor n/2 \rfloor-a] \varepsilon_{a}}
\end{align*}
and we thus need to show that 
\begin{align*}
%\label{error2}
\sum_{a=1}^{\lfloor n/2 \rfloor}{\Pr[X=\lfloor n/2 \rfloor-a] \varepsilon_{a}} & = 1/2 \pm o(1)\,.
\end{align*}
Again similarly as above we bound $\varepsilon_a$ from below
\begin{align*}
\varepsilon_a
& = 
n \left( 
\sum_{i=\lfloor n/2 \rfloor-a+1}^{\lfloor n/2 \rfloor}{1/i} - \sum_{i=\lceil n/2 \rceil+1}^{\lceil n/2 \rceil+a}{1/i}
\right)
 = 
n \left( 
\sum_{i=0}^{a-1}{\frac{1}{\lfloor n/2 \rfloor-i} - \frac{1}{\lceil n/2 \rceil+1+i} }
\right)
\\
& =
n \left( 
\sum_{i=0}^{a-1}{\frac{2i+2}{(\lceil n/2 \rceil-(1+i))(\lceil n/2 \rceil+1+i)} }
\right)
\geq 
n \frac{a^2+a}{(\lceil n/2 \rceil)^2-1}
=
\frac{4(a^2+a)}{n+2-3/n}
\end{align*}
and from above
\begin{align*}
\varepsilon_a 
\leq
\frac{4n(a^2+a)}{n^2-4a^2}\,.
\end{align*}

The variance equals again $n/4$ and can be written as
\begin{align*}
\Var[X] 
& = 
\sum_{a=0}^{\lfloor n/2 \rfloor}{2 \Pr[X=\lfloor n/2 \rfloor-a] (a+1/2)^2}
\\
& = 
1/2+
\sum_{a=0}^{\lfloor n/2 \rfloor}{2 \Pr[X=\lfloor n/2 \rfloor-a] (a^2+a)}
\\
& = 
1/2+
\sum_{a=1}^{\lfloor n/2 \rfloor}{2 \Pr[X=\lfloor n/2 \rfloor-a] (a^2+a)}
\,.
\end{align*}

Thus,
\begin{align*}
\sum_{a=1}^{\lfloor n/2 \rfloor}{\Pr[X=\lfloor n/2 \rfloor-a] \varepsilon_{a}}
& \geq 
\sum_{a=1}^{\lfloor n/2 \rfloor}{\Pr[X=\lfloor n/2 \rfloor-a] \frac{4(a^2+a)}{n+2-3/n}}
\\
& =
(\Var[X]-1/2)\frac{2}{n+2-3/n}
= 
1/2 - o(1)
\end{align*}
and
\begin{align*}
\sum_{a=1}^{\lfloor n/2 \rfloor}{\Pr[X=\lfloor n/2 \rfloor-a] \varepsilon_{a}}
& \leq  
\sum_{a=1}^{\lfloor n/2 \rfloor}{\Pr[X=\lfloor n/2 \rfloor-a]\frac{4n(a^2+a)}{n^2-4a^2}}\,.
\end{align*}
Splitting the sum as in the proof for even values of $n$ shows the desired inequality
\begin{align*}
\sum_{a=1}^{\lfloor n/2 \rfloor}{\Pr[X=\lfloor n/2 \rfloor-a] \varepsilon_{a}}
& \leq  
1/2 +o(1)\,.
\end{align*}

\qed

\noindent\textbf{Acknowledgements.}
Carola Doerr is supported by a postdoctoral research fellowship of the Alexander von Humboldt Foundation and by the Agence Nationale de la Recherche under the project ANR-09-JCJC-0067-01.

}%sloppy
%\bibliographystyle{abbrv}
%\bibliography{references_short}

\end{document}